\documentstyle[amstex,12pt]{article}

\input tcilatex

\begin{document}

\author{Marius I. Piso \\
Institute of Gravitation and Space Sciences\\
Gravitational Researches Laboratory\\
21 Ana Ipatescu bd., 71111 Bucharest, Romania}
\title{Simplicial Euclidean Relativistic Lagrangian }
\date{ }
\maketitle

\begin{abstract}
The paths on the {\bf R$^3$} real Euclidean manifold are defined as
2-dimensional simplicial strips. The Lagrangian of a moving mass is
proportional to the width of the path. The special relativistic form of the
Lagrangian is recovered in the continuum limit, without relativistic Lorenz
invariance considerations\QQfnmark{%
This work has been partially supported on the Romanian Space Agency 1013C
contract.}{}.
\end{abstract}

\section{Introduction}

Standard approaches to discrete space-time consist in the description of the
real space-time manifold as a geometrical lattice \cite{Yamamoto}, \cite%
{Macrae}, \cite{Lehto}or in the space-time approximation by a piecewise
linear space \cite{Regge}, \cite{Ruth}. In this paper, we intend to
introduce a method of discretization which is not performed upon the
space-time manifold, but on the paths, which are defined as 2-dimensional
simplicial strips embedded in an Euclidean space. Preliminary results were
given in \cite{Piso}, \cite{Piso2}.

One of the physical reasons in order to introduce the space-time
discreteness is given by the experimental fact that the speed of light $c$
is maximal and constant; it is easy to explain this by introducing
space-time characteristic constants: an elementary length $\lambda $ and an
elementary time $\tau $. If $\lambda $ and $\tau $ are of Planck dimensions (%
$\lambda _P=(\hbar G/c^3)^{1/2}\simeq 1.616\times 10^{-35}$m and $\tau
_P=\lambda _P/c\simeq 5.391\times 10^{-44}$s) \cite{MTW}, it is generally
accepted that no important influences occur on the continuum space-time
concept at the usual scales of physical phenomena. In the sequel, we
consider $\lambda =\tau =1$, if no other dimensional considerations are
given.

Usual geometrical lattices do not present isotropy, i.e. some favoured
directions are kept in the continuum limit. To circumvent this failure, we
shall make use of a isotropic lattice, which covers the real manifold. Then,
we introduce the simplicial paths. In the final part of the paper, a
connection with the continuum special relativity is presented.

\section{Simplicial paths}

Let ${\bf R}^{3}$ the real linear space endowed with the Euclidean metric.
It is almost obvious that if we accept an elementary length and we need
rotational invariance, the elementary movements should cover the set of all
the possible directions and senses in ${\bf R}^{3}$; this set is isomorphic
to $S^{2}\subset {\bf R}^{3}$ - the 2-sphere. In order to avoid confusions,
we will denote by ${\bf x}$ the vectors of ${\bf R}^{3}$ and by $x$ the
points of the corresponding affine space. Let $\left\vert {\bf x}\right\vert 
$ the length of ${\bf x}$. Let $S^{2}(x_{0})$ the unit length radius sphere
with the center in $x_{0}\in {\bf R}^{3}$. Let ${\bf Z}$ the ring of
integers.\bigskip\ 

\begin{theorem}
There exists at least one embedding of ${\bf R}^3$ in the free ${\bf Z}$%
-module generated by $S^2(x_0)$.$\Box $
\end{theorem}

{\em Proof}. In order to prove the theorem, we show that each ${\bf x}\in 
{\bf R}^3$ has a realization in the ${\bf Z}$-module. Let $a_i\in S^2(x_0)$.
Let ${\bf a}_i\in {\bf R}^3$ the real unit length vectors with the
directions given by $a_j$. An element $\overline{{\bf x}}$ of the free ${\bf %
Z}$-module generated by $S^2(x_0)$ is identified with a real vector ${\bf x}$
as: 
\begin{equation}  \label{1}
\overline{{\bf x}}=\sum_iz_i{\bf a}_i\equiv {\bf x},\quad z_i\in {\bf Z}
\end{equation}
This means that elements of ${\bf R}^3$ could be generated by combinations
of unit length real translations ${\bf a}_i,\left| {\bf a}_i\right| =1$,
with integer coefficients . Let $\forall {\bf x}\in {\bf R}^3$; it is always
possible to explicit the length of ${\bf x}$ as $\left| {\bf x}\right|
=n+\epsilon \,$, where $n\in {\bf Z}^{+}$ (natural number) and $\epsilon \in
[0,1)$. Let ${\bf a}_x={\bf x/}\left| {\bf x}\right| $ the corresponding
unit vector of $S^2(x_0)$ possessing the same direction as ${\bf x}$ . If $%
{\bf e}=\epsilon \cdot {\bf a}_x$ , there always exist two unit vectors $%
{\bf a_\epsilon }$ and ${\bf a_\epsilon ^{\prime }}$ as ${\bf e}={\bf %
a_\epsilon }+{\bf a_\epsilon ^{\prime }}$. In this way: 
\begin{equation}  \label{2}
{\bf x}=n\cdot {\bf a}_x+{\bf e}=n\cdot {\bf a}_x+{\bf a_\epsilon }+{\bf %
a_\epsilon ^{\prime }\;\;},\quad \quad \forall {\bf x}\in {\bf R}^3\quad 
\text{Q.E.D.}
\end{equation}

\begin{corollary}
Let $x,x^{\prime }\in {\bf R}^3$. It is always possible to connect the
points $x$ and $x^{\prime }$ with at least one combination of ${\bf a}_i$
vectors of the form \ref{1}.$\Box $
\end{corollary}

{\em Proof}. Let ${\bf x}={\bf vect(}x,x^{\prime })\in {\bf R}^3$ the vector
defined by the points $x$ and $x^{\prime }$, then see \ref{2} Q.E.D.

\bigskip\ Let us consider that a particle is confined in a space zone of
Planck dimensions; it seems natural that the simplest geometrical model of
the zone is the interior of a sphere $S^2(x)$ of Planck radius $\lambda _P$.
If the particle is considered a point mass, the most probable movement due
to symmetry considerations is the radial one: the particle ''jumps'' from
the center of the sphere to the surface. If we take $\lambda _P=1$, we may
consider that the point particle ''moves'' acted by unit length translations 
${\bf a}_i$; we will call the resultant trajectory a {\em d-path} (d from
discrete).

\begin{definition}
A {\bf d-path} is the ordered set $d_{0N}=\left\{ x_0,x_1,..x_N\right\} $ of
points as: 
\begin{equation}  \label{3}
{\bf vect}(x_0,x_1)={\bf a}_1;\;{\bf vect}(x_1,x_2)={\bf a}_2;..{\bf vect}%
(x_{N-1},x_N)={\bf a}_N\;,{\bf a}_k\in {\bf R}^3,\left| {\bf a}_k\right|
=1\;.
\end{equation}
\end{definition}

In order to describe a movement, we need some topological ingredients to get
the continuity. For an example, if we join two-by-two the consecutive points 
$x_k,x_{k+1}$ with segments, we get a 1-dimensional simplicial complex made
of unit length edges. The polyhedra attached to such 1-dimensional
simplicial complexes are called {\em edge-paths}; their properties are known
in algebraic topology \cite{Maunder}. However, if we describe the
trajectories by the mean of edge-paths, the immediate conclusion is that
movement could develop only with the speed of light. In order to remove this
difficulty, the following constructions are introduced:

\begin{definition}
Let $d$ a d-path defined by the set of points $\left\{ x_0,x_1,..x_N\right\} 
$. Let $a_k=\overline{x_{k-1}x_k}$ the edges defined by the points $x_{k-1}$
and $x_k$. Let $\sigma _k$ the 2-dimensional simplex (isosceles triangle)
defined by the segments $a_k$ and $a_{k+1}$. The {\bf D-path} $D$ is the
polyhedron $\left| \Sigma \right| $ attached to the 2-dimensional simplicial
complex $\Sigma $, as follows: 
\begin{equation}  \label{4}
\Sigma =\left\{ x_0,x_1,..x_N\right\} \cup \left\{ a_1,a_2,..a_k\right\}
\cup \left\{ \sigma _1,\sigma _2,..\sigma _{k-1}\right\} \quad \quad D\equiv
\left| \Sigma \right| .
\end{equation}
\end{definition}

\begin{definition}
The {\bf speed} attached to the simplex $\sigma _k$ component of $D$ is the
vector ${\bf v}_k$ defined by the baricentres $\hat x_{k-1}$ and $\hat x_k$
of the segments $a_k$ and $a_{k+1}$ which generate the simplex $\sigma _k$: 
\begin{equation}  \label{5}
{\bf v}_k={\bf vect}(\hat x_{k-1},\hat x_k)\;.
\end{equation}
\end{definition}

It is easy to note that $\left| {\bf v}_k\right| \in [0,1]$ as in the
special relativity theory. We mention that the procedure of baricentric
subdivision is known in algebraic topology as a method to construct the
derivative of a simplicial complex \cite{Maunder}; in this way, the edge $%
v_k=\overline{\hat x_{k-1}\hat x_k}$ of our construction may be regarded as
the derivative of $\sigma _k$, or the derivative of $D$ at time $k$.

\bigskip\ The standard procedure to describe the movement of a point-like
particle on the ${\bf R}^3$ manifold is to introduce the paths described by
the particle, which are orbits of the action of some real one-parameter
subgroup $G$ on ${\bf R}^3$, in such a manner that: 
\begin{equation}  \label{6}
[0,T]\overset{G}{\longmapsto }{\bf x}(t)|_0^T
\end{equation}
with ${\bf \dot x}(t)\equiv d/dt[{\bf x}(t)]$ the speed. In the present
approach, $G$ is replaced by a discrete one- parameter subgroup ${\cal G}$,
i.e.: 
\begin{equation}  \label{7}
[0,1,...N]\overset{{\cal G}}{\longmapsto }\overset{d-path}{\left\{
x_0,x_1,..x_N\right\} }\longmapsto \overset{D-path}{\left\{ \sigma _1,\sigma
_2,..\sigma _{N-1}\right\} }
\end{equation}
(the lose of the end points of the discrete time interval is not significant
for this approach). By means of the baricentric subdivision, we attached the
speed ${\bf v}_k$ (see \ref{5}) to each simplex $\sigma _k$ component of the 
{\em D-path} $\left\{ \sigma _1,\sigma _2,..\sigma _{N-1}\right\} $ in such
a manner that ${\bf v}(\sigma _k)={\bf v}(k)$.

\bigskip\ It is important to notice the difference between a simple discrete
path, which is an ordered set of disconnected points of ${\bf R}^3$, and a 
{\em D-path}, which is an ordered set of geometric 2-simplexes connected by
edges, whose topology is the one induced by the attached polyhedron (see 
\cite{Maunder}).

It is easy to observe the correspondence with the continuum mechanics: for
large speed ($\left| v_k\right| \rightarrow 1$), the {\em D-path} goes to a 
{\em d-path}, which is 1-dimensional; for small speed, the {\em D-path}
becomes a strip of width $\simeq \lambda $ and the speed goes to a vector
tangent to the median line of the strip, line which identifies itself with
the classical trajectory. In this sense, we shall define the continuum limit
as follows.

\begin{definition}
Let $D$ a D-path. The {\bf 1-dimensional path} attached to $D$ is the
polyhedron $\left| V\right| $ attached to the 1-dimensional simplicial
complex $V$ defined by the baricenters $\hat x_k$ of the edges of the
corresponding d-path: 
\begin{equation}  \label{8}
V=\left\{ \hat x_0,\hat x_1...\hat x_N\right\} \cup \left\{
v_1,v_2,...v_{N-1}\right\} ;\quad v_k=\overline{\hat x_{k-1}\hat x_k}\;\;.
\end{equation}
\end{definition}

Within the upper definition, the 1-dimensional path is a triangulation \cite%
{Maunder} of a classical path, made of edges of length $l_k=\left|
v_k\right| \in [0,1]$ or, with dimensional considerations, $l_k\in
[0,\lambda ]$. The speed attached to the edge is ${\bf v}_k$ (see \ref{5}).
The coordinate ${\bf x}_k$ of the edge may be considered the position of
either $\hat x_k$, or the baricenter of $v_k$, or even the baricenter of $%
\sigma _k$. The time coordinate is $k\in \{0,1...N\}$ or, with dimensional
considerations, $\tau \cdot k$. In this way, we may write:%
\[
\triangle {\bf x}(k)=c\cdot {\bf v}_k\cdot \tau \quad \longrightarrow \quad d%
{\bf x}={\bf v}(t)\cdot dt 
\]
\[
\triangle l(k)=c\cdot v_k\cdot \tau \quad \longrightarrow \quad ds=v(t)\cdot
dt 
\]
\begin{equation}  \label{9}
l_{0N}\equiv \text{ the length of the path }=\lambda \cdot
\sum_{k=1}^Nv_k\quad \longrightarrow \quad
\int_{x_0}^{x_N}ds=\int_{t_0}^{t_N}v(t)dt
\end{equation}
We regain in this way the finite differences methods and go to the continuum
limit by means of standard procedures.

\section{The Lagrangian}

{}From the preceding section, we reached to the concept of {\em D-path},
which contains by itself sufficient parameters in order to define the
movement: {\em position} ${\bf x}_k$ and {\em speed} ${\bf v}_k$. We elude
the concept of point mass, if we consider that the mass is distributed on
the 2-simplex $\sigma _k$. In this case, in a good approximation, ${\bf x}_k$
may be identified with ${\bf \hat x}(\sigma _k)$ - the position of the
baricenter of the simplex $\sigma _k$.

In order to construct the Lagrangian ${\em L}({\bf x}_k,{\bf v}_k)$ for a
mass {\em m$_0$} moving on a {\em D-path}, we shall start from the following
physical considerations:\medskip\ 

\begin{quote}
i) ${\em L}$ must be invariant under the Euclidean rotations and
translations, in order to preserve the fundamental properties of
movement;\medskip\ 

ii) ${\em L}({\bf x}_k,{\bf v}_k)$ must be a specific component of the {\em %
D-path}, which contains sufficient variables to describe the
movement.\medskip\ 
\end{quote}

According to the geometric construction of the {\em D-path}, the magnitude $%
v_k$ of the speed ${\bf v_k}$ on $\sigma _k$ may be written as follows: 
\begin{equation}  \label{10}
v_k=\cos \frac{\alpha _k}2\quad ,\quad \alpha _k=\pi -\arccos ({\bf a}%
_{k-1}\cdot {\bf a}_k)
\end{equation}
where $\alpha _k$ is the angle between the edges $a_{k-1}$ and $a_k$ which
determine the simplex $\sigma _k$ (see \ref{4}). In terms of speed $v_k$
(see \ref{8}), the length of the height line (normal to $v_k$) of the
geometrical simplex $\sigma _k$ becomes: 
\begin{equation}  \label{11}
h_k=\sin \frac{\alpha _k}2=\sqrt{1-v_k^2}
\end{equation}
proportional to the Lorenz square root.

In the continuum limit ($\lambda ,\tau \rightarrow 0$ as $\lambda /\tau =c$%
), $h_k$ identifies itself with the transversal width of the strip defined
by the {\em D-path} at the position $x_k$. Using adequate units \ref{9} (see
also section I), we get for $h_k$: 
\begin{equation}  \label{12}
\frac{h_k}\lambda =\sqrt{1-\frac{v_k^2}{c^2}}\Longleftrightarrow \frac{h(k)}%
\lambda =\sqrt{1-\frac{v^2(k)}{c^2}}\overset{\text{continuum limit}}{%
\longrightarrow }\sqrt{1-\frac{v^2(t)}{c^2}}
\end{equation}
The relativistic Lagrangian for a point particle of rest mass ${\em m}_0$ is
defined in continuum relativistic mechanics as \cite{MTW}: 
\begin{equation}  \label{13}
{\em L}(v)=-m_0c^2\sqrt{1-\frac{v^2(t)}{c^2}}
\end{equation}
After dimensional considerations we get from \ref{12} and \ref{13} the
following:\bigskip\ 

\begin{definition}
The {\bf Lagrangian} $L(v_k)$ on a {\em D-path} is defined as: 
\begin{equation}  \label{14}
L(v_k)\equiv -m_0c^2\frac{h_k}\lambda
\end{equation}
proportional to the {\em height} $h_k$ of the simplex $\sigma _k$.\bigskip\ 
\end{definition}

In the continuum limit, we get:

\begin{equation}  \label{15}
{\em L}(v_k)=-m_0c^2\frac{h_k}\lambda \longrightarrow -m_0c^2\frac{h(t)}%
\lambda =-m_0c^2\sqrt{1-\frac{v^2(t)}{c^2}}
\end{equation}
that the simplicial Lagrangian \ref{14} goes to the standard relativistic
Lagrangian. This is surprising, because no relativistic invariant has been
involved in the construction of the model. It is also interesting to note
that the Lagrangian \ref{14} is independent on the magnitude of the
elementary length $\lambda $.

\section{Conclusions}

The starting point of our analysis is the idea that the speed of light is
maximal and constant due to some discrete properties of the space-time. In
the present approach, we did not discretize the space-time manifold, but the
paths. First, we took as base manifold the ${\bf R}^{3}$ Euclidean space and
we considered that an elementary discrete movement is a set of sequentially
equidistant points of ${\bf R}^{3}$. In order to emphasize the rotational
invariance, we showed in the second section that these sets are elements of
the free ${\bf Z}$-module generated by the $S^{2}\subset {\bf R}^{3}$ unit
radius sphere and we proved that all ${\bf R}^{3}$ is covered by means of
this procedure. In order to give a topology (to introduce the continuity) on
the discrete sets of points, we defined the {\em D-paths} as 2-dimensional
simplicial complexes having as vertices these sets. By means of a
baricentric subdivision, we attached to each simplex component of the D-path
a {\em speed} whose magnitude varies between $0$ and $1$. In the third
section, we defined the Lagrangian of a moving mass (distributed on a
2-simplex) as being proportional to the transversal width of the simplicial
strip. In the continuum limit, the Lagrangian becomes the correspondent
relativistic one for the point particle. In this way, we are able to claim
that the basic special relativistic properties of movement may be derived as
a consequence of the local discreteness of the space-time manifold.

\newpage\

\QQfntext{0}{
This work has been partially supported on the Romanian Space Agency 1013C
contract.}
\end{document}